\documentclass[prl,showpacs,preprintnumbers,amsmath,amssymb,twocolumn]{revtex4}

\usepackage{graphicx}
\usepackage{dcolumn}
\usepackage{bm}
\usepackage{color}
\usepackage{layout}
\usepackage{verbatim}

\newcommand{\beq}{\begin{equation}}
\newcommand{\eeq}{\end{equation}}

\newcommand{\nn}{n_{\pi}}
\newcommand{\TT}{\tau_{\text{D}}}
\newcommand{\GG}{\gamma_{\rm{e}}}

\begin{document}

\title{Scaling of Dynamical Decoupling for Spin Qubits}

\date{\today}

\author{J.~Medford$^{1}$}
\author{\L.~Cywi\'{n}ski$^2$}
\author{C.~Barthel$^{1}$}
\author{C.~M.~Marcus$^1$}
\author{M.~P.~Hanson$^3$}
\author{A.~C.~Gossard$^3$}
\affiliation{$^1$Department of Physics, Harvard University, Cambridge, Massachusetts 02138, USA\\
$^2$Institute of Physics, Polish Academy of Sciences, Al.~Lotnik\'{o}w 32/46, PL 02-668 Warszawa, Poland\\
$^3$Materials Department, University of California, Santa Barbara, California 93106, USA
}

\begin{abstract}
We investigate scaling of coherence time, $T_2$, with the number of $\pi$-pulses, $n_{\pi}$, in a singlet-triplet spin qubit using Carr-Purcell-Meiboom-Gill (CPMG) and concatenated dynamical decoupling (CDD) pulse sequences. For an even numbers of CPMG pulses, we find a power law, $T_{2} \propto (n_{\pi})^{\GG}$, with $\GG = 0.72\pm0.01$, essentially independent of the envelope function used to extract $T_2$. From this surprisingly robust value, a power-law model of the noise spectrum of the environment, $S(\omega) \sim \omega^{-\beta}$, yields $\beta = \GG/(1-\GG) = 2.6\pm0.1$.  Model values for $T_2(n_{\pi})$ using $\beta = 2.6$ for CPMG with both even and odd $n_{\pi}$ up to 32 and CDD orders 3 through 6 compare very well with experiment.
\end{abstract}


\maketitle

A variety of solid state systems are emerging as effective platforms for studying decoherence and entanglement in controlled quantum systems \cite{Bluhm_NP11,Barthel_PRL10,Bylander_NP11,deLange_Science10}. 
Among them, quantum-dot-based spin qubits have recently achieved sufficient control and long coherence times \cite{Bluhm_NP11,Barthel_PRL10} that new information about the noise environment of the qubit can be extracted, complementing related work in nitrogen-vacancy centers in diamond \cite{deLange_Science10}, superconducting qubits \cite{Bylander_NP11}, trapped ions \cite{Biercuk_Nature09}, and neutral atoms \cite{Sagi_PRL10}.  

Dynamical decoupling in the form of a sequence of $\pi$-pulses \cite{Viola_PRA98,Uhrig_PRL07,Khodjasteh_PRA07,Khodjasteh_PRA11} functions as a high-pass filter, thus providing information about the spectral content of environmental noise \cite{deSousa_TAP09,Cywinski_PRB08,Biercuk_Nature09,deLange_Science10,Biercuk_JPB11,Bylander_NP11,Biercuk_PRB11}. 
For spin qubits, the effectiveness of various decoupling schemes at mitigating dephasing due to nuclear bath dynamics has been well studied theoretically \cite{Yao_PRB06,Witzel_PRB06,Witzel_PRL07,Cywinski_PRB09,Neder_PRB11}. Much less is known about mitigating the effects of charge noise, which couples to the qubit via gate dependent exchange interaction and through spatially varying Overhauser fields \cite{Bluhm_NP11}. When the decoherence time, $T_{2}$, is short compared to the energy relaxation time, $T_{1}$,---which is the case in this study---both the envelope of the coherence decay as well as the dependence of $T_{2}$ on the number of  $\pi$-pulses, $\nn$, depend on the spectral density of the environment, $S(\omega)$. Knowledge of $S(\omega)$ inferred from such measurements can in turn be used to design optimal decoupling sequences \cite{Cywinski_PRB08,Gordon_PRL08,Biercuk_Nature09,Pan_PRA10,Ajoy_PRA11}.

\begin{figure}[b]
\includegraphics[width = 3
 in]{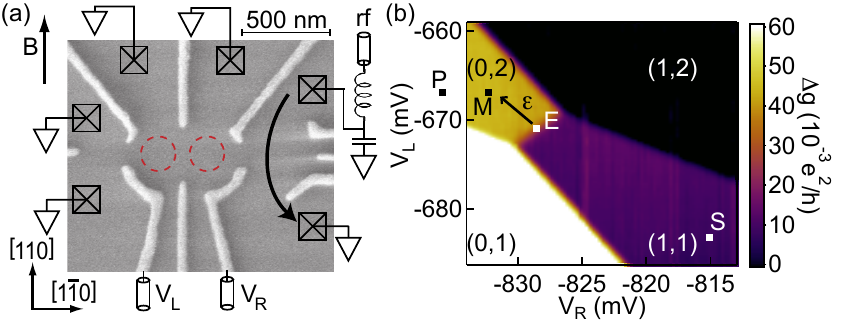}
\caption{\label{FigDevice}~(Color online)~(a)~Micrograph of lithographically identical device with dot locations depicted. Gate voltages, $V_{\rm{R(L)}}$, set the charge occupancy  of right (left) dot as well as the detuning of the qubit. An rf-sensor quantum dot is indicated on the right.
(b) Double dot charge state mapped onto dc conductance change, $\Delta g$, with lettered pulse sequence gate voltages. Detuning axis is orthogonal to the (0,2)-(1,1)  charge degeneracy through points E, S, and M. }
\end{figure}

In this Letter, we  investigate  scaling of $T_{2}$ with the number of $\pi$-pulses for Carr-Purcell-Meiboom-Gill (CPMG) and concatenated dynamical decoupling (CDD) sequences in a GaAs two-electron singlet-triple qubit [Fig.~1(a)]. The coherence envelope is reasonably well described by the form $\exp(-(\TT/T_2)^\alpha)$, where $\TT$ is the time during which $\pi$-pulses are applied [Fig.~2(b)]. It is difficult, however, to accurately determine $\alpha$ by directly fitting to this form. In contrast, we find that the scaling relation $T_2\sim(\nn)^{\gamma}$ very accurately describes the data irrespective of the value of $\alpha$ used to extract $T_{2}$. The resulting $\gamma$ can then be related to $\alpha$ and other quantities of interest within specific noise models. For CPMG with even $\nn$, the scaling relation $T_2 \propto (\nn)^{\GG}$ yields $\GG = 0.72\pm0.01$, using $T_{2}$ values extracted using any $\alpha$ in the range 2 to 5. A model of dephasing due to a power-law spectrum of classical noise, $S(\omega)\! \sim \! \omega^{-\beta}$, leads to a scaling relation in the number of $\pi$-pulses, with the exponent of the power law, $\beta$, related to the scaling exponent by the simple relation $\beta=\GG/(1-\GG)$. For the present experiment, $\GG = 0.72$ thus yields $\beta = 2.6$. Further support for a power-law form for $S(\omega)$ is found by comparing experimental and theoretical dependences $T_{2}(n_{\pi})$ for CPMG with both even and odd $n_{\pi}$ as well as CDD pulse sequences. This model also gives the simple relation $\alpha = \beta + 1$ connecting the noise spectrum and the decoherence envelope exponent. The resulting value, $\alpha = 3.6\pm0.1$ is thus determined with considerably greater accuracy than can be obtained from direct fits to the coherence envelope data.

The lateral double quantum dot investigated was defined by Ti/Au depletion gates patterned using electron beam lithography on a GaAs/Al$_{0.3}$Ga$_{0.7}$As heterostructure with a two dimensional electron gas (density $2 \times 10^{15}$ m$^{-2}$, mobility 20 m$^2/$V s) 100 nm below the surface. Measurements were performed in a dilution refrigerator with an electron temperature $T_e\!\!\sim\!\!150$ mK. The double quantum dot is operated as a spin qubit by first depleting the quantum dots to the last two electrons, then manipulating the charge occupancy of the two dots with high bandwidth plunger gates V$_{\rm{L}}$ and V$_{\rm{R}}$ along a detuning axis $\epsilon$ [Fig.~1(b)]. In this work, the charge occupancy was manipulated between states (0,2) and (1,1), where (N$_{\rm{L}}$,N$_{\rm{R}}$) represent the charge in the left and right dots. Charge occupancy was determined by the conductance change, $\Delta g$, through a proximal sensor quantum dot, which in turn modulated the reflection coefficient of the radio-frequency (rf) readout circuit \cite{Reilly_APL07,Barthel_PRB10}. 

The logical spin qubit subspace is spanned by the singlet $(S= (|\!\!\uparrow\downarrow\rangle - |\!\!\downarrow\uparrow\rangle)/\sqrt{2}))$ and the $m=0$ triplet $(T_0= (|\!\!\uparrow\downarrow\rangle+|\!\!\downarrow\uparrow\rangle)/\sqrt{2}))$ states of two electrons. The $m=\pm1$ triplet states were split off by a 750 mT magnetic field applied in the plane of the electron gas, perpendicular to the dot connection axis. A (0,2) singlet 
was prepared at point P, off the detuning axis, through rapid relaxation to the ground state, then moved to the separation point S in (1,1). Uncorrelated Overhauser fields in the two dots create an evolving Zeeman gradient, $\Delta B_{z}$,  that drives transitions between $S$ and $T_0$.  Single-shot readout was performed by moving  to point M, where $S$ can tunnel to (0,2) while $T_0$ remains in (1,1). The reflectometer signal was integrated for 600 ns per shot, averaged over $10^4$ shots, and compared to voltage values corresponding to $S$ and $T_0$ outcomes \cite{Barthel_PRL09}, yielding $P_S(\TT)$, the probability of singlet return.

\begin{figure}[t]
\includegraphics[width = 3.2 in]{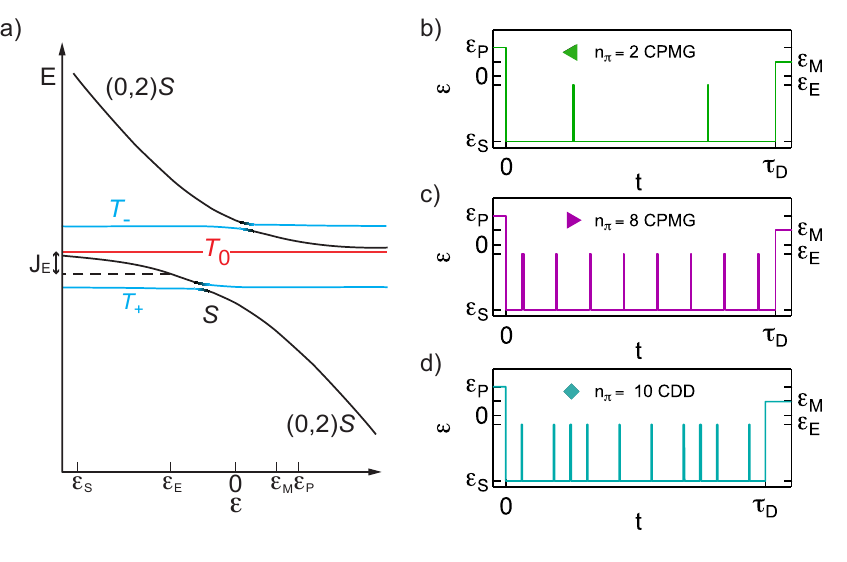}
\caption{
\label{FigPulse}
(Color online)~(a)~Energy level diagram along the detuning axis $\epsilon$. The (0,2) singlet was prepared at $\epsilon_{\rm{P}}$, followed by separation to $\epsilon_{\rm{S}}$. $\pi$-pulses are performed at $\epsilon_{\rm{E}}$, allowing for subsequent rephasing at  $\epsilon_{\rm{S}}$. Single-shot readout at $\epsilon_{\rm{M}}$ using a proximal sensor dot. The exchange energy, $J_{\rm{E}}$, that drives the $\pi$-pulses at $\epsilon_{\rm{E}}$ is indicated with a dashed line. (b-d) Schematics of detunings during CPMG and CDD pulse sequences with detuning points on the vertical axis. }\end{figure}

Coherence lost due to (thermally driven) evolution of $\Delta B_{z}$ can be partially restored using a Hahn echo by pulsing at time $\tau_{D}/2$ to point E, where the exchange splitting between $S$ and $T_0$ drives a $\pi$-rotation about the $\hat{x}$ axis, changing the sign of the acquired phase. Returning to S for an equal time $\tau_D/2$ cancels the phase acquired due to the low-frequency ($\omega \! < \! 2/\tau_{D}$) end of the spectrum of fluctuations of  $\Delta B_{z}$ \cite{deSousa_TAP09,Cywinski_PRB08}.  
Dynamical decoupling using a series of $\pi$-pulses allows efficient removal of more of the low-frequency end of the noise spectrum \cite{deSousa_TAP09,Cywinski_PRB08}. The CPMG sequence \cite{Vandersypen_RMP05}, for example, uses evenly spaced gate pulses from point S to point E with a half interval before the first and after the last $\pi$-pulse [Figs.~\ref{FigPulse}(b,c)].  Concatenated dynamical decoupling \cite{Khodjasteh_PRA07} (CDD) uses nonuniformly spaced pulses to point E, where the $k$-th order sequence is determined recursively from the lower order one, 
with an additional $\pi$-pulse in the center of odd orders [Fig.~\ref{FigPulse}(d)].

\begin{figure}
\centering
\includegraphics[width = 2.5 in]{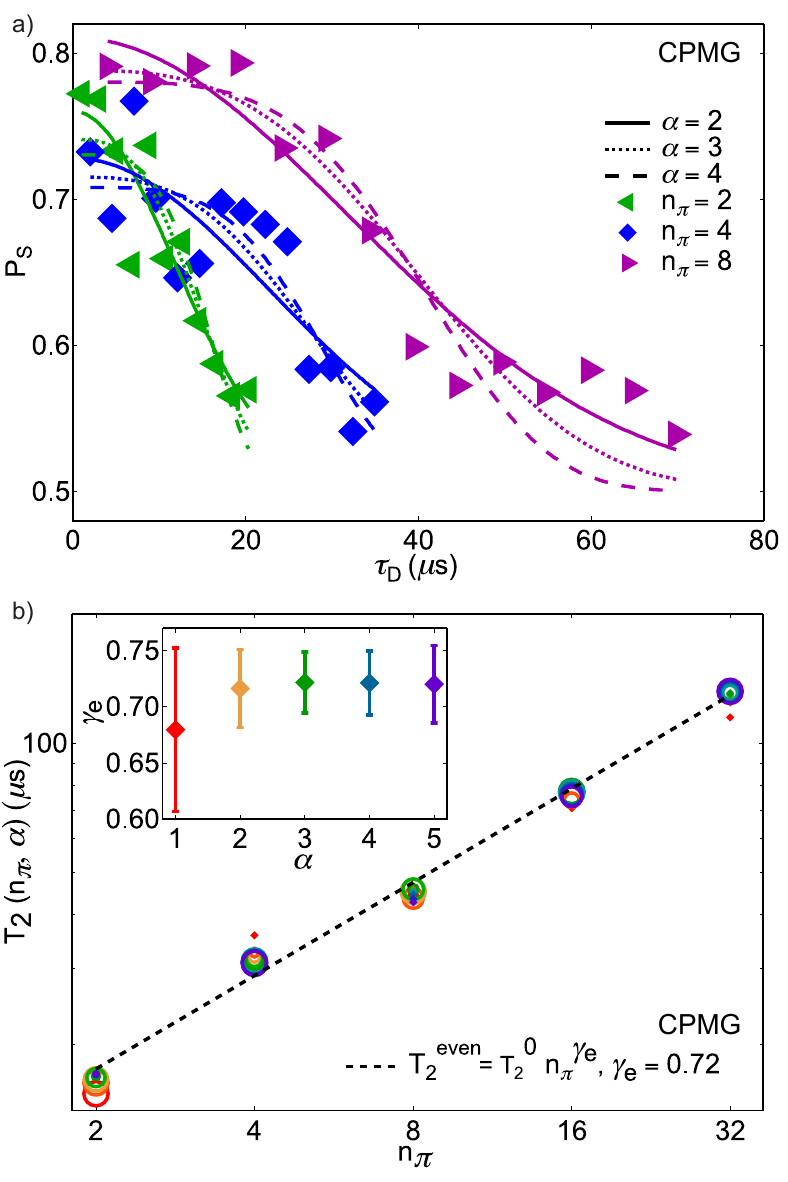}
\caption{
\label{FigCPMG}~(Color online)~(a)~Experimental singlet return probabilities as a function of time for CPMG with $n_{\pi} = 2,4,8$. Fits to $P_S(\TT) = 0.5 + V/2 \exp(- (\TT/T_2)^\alpha)$, with $\alpha$ constrained to 2 (solid curves), 3 (dotted curves) and 4 (dashed curves) \cite{CPMGa2to4params}. It is difficult to determine $\alpha$ from these fits.
(b)~ Extracted $T_2$ for even-$\nn$ CPMG sequences for $\alpha$ constrained to 1, 2, 3, 4, and 5. Circle size proportional to $\chi^2$ goodness of fit of $P_S(\TT)$ in (a). A power-law fit to the form $\ln(T_2^{\rm even}(\alpha)) = \ln(T_2^{0}) + \GG \,\ln(n_{\pi})$, shown for $\alpha$=3 (dashed line) gives $\GG = 0.72$. The fit value $\GG$ depends only weakly on $\alpha$ in the range $2-5$ (inset). The weighted average over $\alpha = 2-5$ yields $\GG = 0.72\pm 0.01$.}
\end{figure}

\begin{figure}[t]
\centering
\includegraphics[width = 2.5 in]{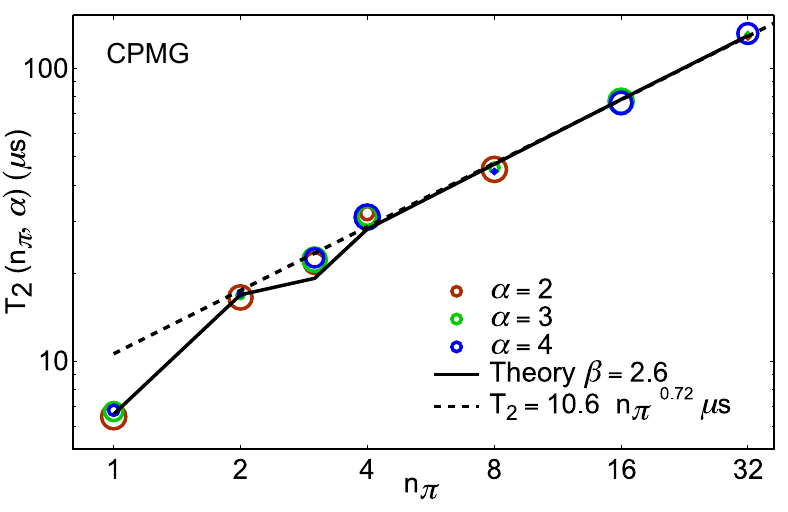}
\caption{
\label{FigFullTheory}~(Color online)~ $T_2$ for all measured $\nn$ for CPMG, extracted using $\alpha =$ 2, 3, and 4 (circles). Circle size for each $\alpha$ proportional to $\chi^{2}$ goodness of fit.  Theory (black solid curve) for $\beta = \GG/(1-\GG)=2.6$ and integration of Eq.~(\ref{eq:chi}) with the CPMG filter functions. Note that Eq.~(\ref{eq:chiCPMG}) captures the even/odd effect quantitatively for small $\nn$.
}
\end{figure}

Singlet return probabilities $P_S(\TT)$ were measured for CPMG sequences with $\nn$= 1, 2, 3, 4, 8, 16, and 32. Fits to $P_S = 0.5 + V/2 \exp[-(\TT/T_2)^\alpha]$, with visibility $V$, $T_{2}$, and $\alpha$ as a fit parameter, were equally good for $\alpha$ between 2 and 4, as seen in Fig.~3(a). For this reason, though $S(\omega)$ is related to $\alpha$, these fits give little information about the spectrum of the environment. Figure 3(b), showing $T_{2}$ as a function of $\nn$ for an even number of CPMG pulses, shows two remarkable features. First, values of $T_{2}$ do not depend on the value of $\alpha$ used in the fits to $P_S(\TT)$. Second, $T_2$ shows a power-law scaling $T_2 = T_2^{0} (\nn)^{\gamma}$ whose power, $\gamma$, also does not depend on the value of $\alpha$ used in the fits.

To model these observations, we consider Gaussian noise affecting the energy splitting of the qubit, which leads to the off-diagonal (in the basis of $|\!\!\uparrow\downarrow\rangle$ and $|\!\!\downarrow\uparrow\rangle$) elements of the qubit density matrix decaying as $\exp[ -\chi(\TT) ]$, where
\beq
\chi(\tau_{D}) = \int_{0}^{\infty} \frac{\text{d}\omega}{\pi} S(\omega) \frac{F(\omega \TT)}{\omega^2} \,\, ,  \label{eq:chi}
\eeq
with  $F(\omega\TT)$ being the filter function determined by the sequence of $\pi$-pulses driving the qubit. For CPMG sequence $F(z) \! < \! (z/2\nn)^{4}$ for $z\! < \! 2\nn$, i.e.~$F(z)$ strongly suppresses the low-frequency noise, while for large $z$ and $\nn$ the filter function can be approximated \cite{Cywinski_PRB08} by a periodic train (with period $z_{p} \! = \! 2\pi\nn$) of square peaks of height $h \! \approx \! 2\nn^2$ and width $\Delta z \! \approx \! z^{2}_{p}/\pi h \! \ll \! z_{p}$.

We find that the value of $\GG$ and the presence of an even-odd effect (EOE) in $\nn$ (i.e.~$\GG \! \neq \! \gamma_{\rm{o}}$) act as discriminators for several classes of $S(\omega)$: $(i)$ The case of $0<\GG\leq 2/3$ and absent EOE is compatible with a model of $S(\omega)\sim \omega^{-\beta}$ (over a range of $\omega$ roughly bounded by the minimal and maximal values of $\nn/\TT$), with $0 \! <\! \beta \! \leq \! 2$. In this case $\GG = \beta/(1+\beta)$ and  $\alpha \! =\! \beta+1$. One example is the case of Ornstein-Uhlenbeck noise \cite{Dobrovitski_PRL09} (having Lorentzian $S(\omega)$ with $\omega^{-2}$ tail typically dominating the decoherence under dynamical decoupling), 
where $\gamma \! = \! 2/3$ was confirmed by experiments on the NV center \cite{deLange_Science10}. $(ii)$ 
When an EOE is present, $\GG = 2/3$ suggests a hard cutoff in $S(\omega)$ at $\omega_c \! < \! 2/T_{2}$, in which case  $\gamma_{o} \! = \! 1$. $(iii)$ For $\omega_{c}\! > \! 2/T_{2}$, i.e.~for larger $\omega_{c}$ or larger $\nn$ (leading to longer $T_{2}$), the EOE disappears and $\gamma$ tends to $1$ \cite{Biercuk_PRB11}.
$(iv)$ Finally, the presence of the EOE and $2/3\!<\!\GG\! < \! 1$ indicate $S(\omega)\sim \omega^{-\beta},$ with $\beta \! > \! 2$. 

Experimentally, we find $\GG \! = \! 0.72$ for even number of CPMG pulses, and $\nn=1$ not along the scaling line, indicating an EOE. We conclude that scenario $(iv)$ applies, namely $S(\omega) \! = \! A^{\beta+1}/\omega^{\beta}$ with $2 \! < \! \beta \! < \! 3$. 
Using Eq.~(\ref{eq:chi}) and the CPMG filter function gives in the large-$\nn$ limit
\beq
\chi(t) \approx (A\TT)^{1+\beta} \left( \frac{a}{\nn^{\beta}} + \frac{b_{e/o}}{\nn^4} \right )\,\, ,  \label{eq:chiCPMG}
\eeq
with $a \! \approx \! \Sigma_{2+\beta}/\pi^2(2\pi)^{\beta}$, where $\Sigma_{\delta} \! = \! \sum_{k=1}^{\infty} (k-\frac{1}{2})^{-\delta}$, and for odd (even) $\nn$ we have $b_{o} \! \approx \! [32\pi(3-\beta)]^{-1}$ ($b_{e}  \approx [128\pi(5-\beta)]^{-1} \approx b_{o}/10$), i.e.,~the $b/\nn^4$ term is negligible for even $\nn$, while it gives a significant correction for small, odd $\nn$. The EOE comes from the difference in the low-$z$ behavior of the CPMG filter functions, which for $z \! < \! 1$ behave as $F(z) \! \sim \! z^{4}/2^{5}\nn^{4} \, (z^{6}/2^{7}\nn^{4})$ for odd (even) $\nn$. For $\beta \! > \! 2$ this leads to different contributions of very low $\omega$ to the integral in Eq.~(\ref{eq:chi}). 
For even $\nn$, we find that $\chi(\TT)$ approximately reduces to $(A\TT)^{1+\beta}a/\nn^\beta$, from which we obtain the $\beta \leq 2$ result of $\GG = \beta/(1+\beta)$ in this case as well.

Assuming this form of $S(\omega)$, fits to the even $\nn$ [Fig.~3(b)] yield $\beta = 2.6$ and $A^{-1}\! = 3.6$ $\mu$s. Using these two parameters we  calculate odd-$\nn$ values for $T_{2}$ by numerically integrating Eq.~(\ref{eq:chi}). As shown in Fig.~\ref{FigFullTheory}, the obtained value of $T_{2}$ is in good agreement with the measured value for $\nn \! = \! 1$ (Hahn echo). 
We note that the large $\nn$ scaling of $T_2 \! \sim \! \nn^{\gamma}$ is due to the behavior of $S(\omega)$ at $\omega  \! \geq \! \pi \nn / T_{2}$, which is $\sim 0.3 (\nn)^{0.28}$ $\mu$s$^{-1}$ here, while the EOE at small $\nn$ is due to behavior at $\omega \! < \!1/T_{2}$, which is $\sim\! 0.15$ $\mu$s$^{-1}$. The consistency between small- and large-$\nn$ data indicates that $S(\omega) \! \sim \! \omega^{-2.6}$ over this range of frequencies (i.e.,~$ \omega/2\pi \sim 10-100$ kHz).
The EOE behavior at low $\nn$ can be fit within scenario $(ii)$ using $S(\omega) \! =\! A^3/\omega^2$ with $A^{-1} \! \sim \! 1$ $\mu$s and  $\omega_{c} \! \sim \! 0.08$ $\mu$s$^{-1}$. However, this scenario crosses over to $(iii)$ for $\nn > 5$,  where $\gamma$ tends to $1$. The resulting large-$\nn$ behavior, $T_{2} \! \sim \! \nn \times 7$ $\mu$s,  departs significantly from the $\nn \! \geq \! 8$ data in Fig.~4.

Using  $S(\omega) \! = \! A^{\beta+1}/\omega^{\beta}$ with parameters $A$ and $\beta$ fixed from the even-$\nn$ CPMG fit, we can calculate the expected dependence of $T_{2}(\nn)$ for the CDD pulse sequence using the known filter functions \cite{Cywinski_PRB08}. For $\nn \! = \! 5$, $10$, and $21$ we get good agreement between the calculated and measured $T_{2}$ [Fig.~5]. For $\nn\!=\!42$, the experimental $T_{2}$ is shorter than predicted by theory, possibly reflecting an accumulation of errors for such a large number of pulses. Note that CDD was shown to be robust to pulse errors \cite{ZhiHui_arXiv10} only in the case of two-axis control (i.e.~when the $\pi$-pulses are about $x$ and $y$ axes alternately) and for a quasi-static bath. 

\begin{figure}
\centering
\includegraphics[width = 2.5 in]{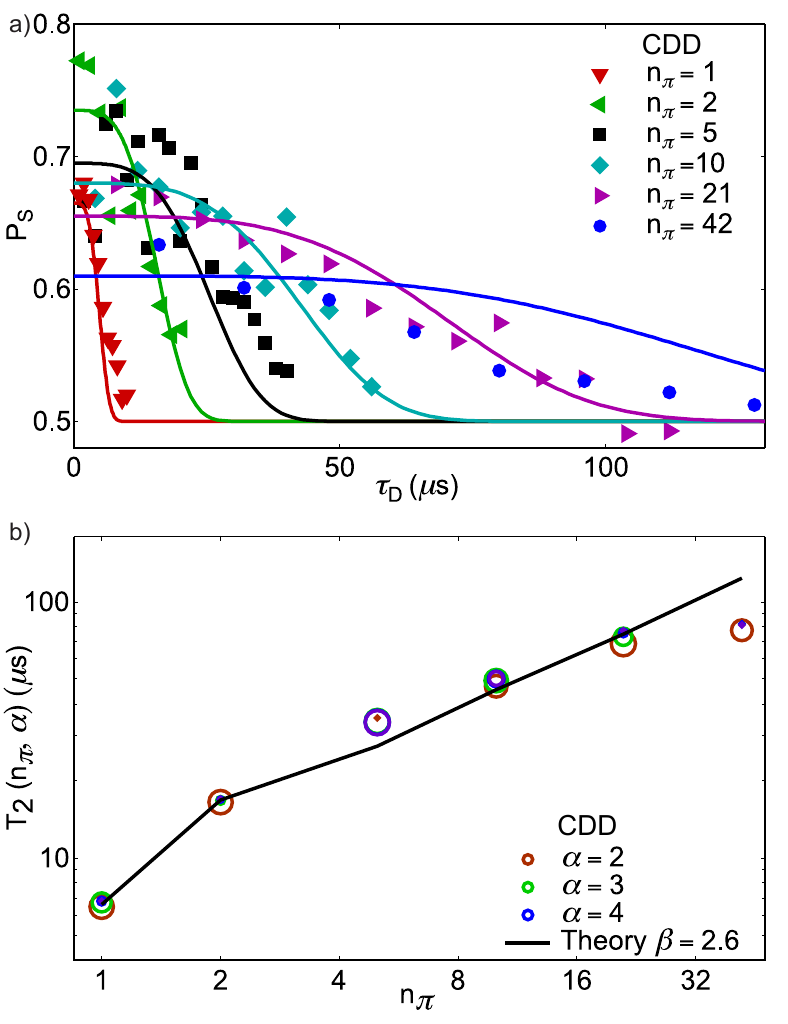}
\caption{
\label{FigCDD}~(Color online)~(a)~Experimental singlet return probabilities for orders 1 through 6 CDD (symbols) along with fits to $P_S(\TT)= 0.5 + V/2 \exp(- (\TT/T_2)^\alpha)$) using $\beta = \GG/(1-\GG)=2.6$ (solid curves).  (b) Extracted $T_{2}$ for using different values of $\alpha$ (circle size proportional to $\chi^2$ measure of goodness of fit). Theory (solid curve) based on $\beta = \GG/(1-\GG)=2.6$ and integration of Eq.~(\ref{eq:chi}) with appropriate CDD filter function \cite{Cywinski_PRB08}.
}
\end{figure}

Summarizing, the measurements of qubit decoherence under dynamical decoupling with the CPMG pulse sequence have been used to reconstruct the crucial features of the spectral density of noise dephasing the qubit. Using the data for even $\nn$ of CPMG we have been able to estimate the magnitude of noise and its functional form. The reconstructed spectral density of noise allows us to calculate the expected decoherence signal for other pulse sequences, and this calculation agrees with the CDD sequence measurements for $\nn$ as well as the odd-$\nn$ CPMG data. We have shown that instead of fitting the exact functional form of the coherence decay function, an analysis of the scaling of the measured $T_{2}$ time with the number of applied pulses allows for a clearer understanding of the system. We cannot say at this point whether the observed value $\gamma_{\rm e} = 0.72$ is characteristic of Overhauser-dominated dephasing in general, or just our particular combination of noise sources.

\begin{acknowledgments}
We acknowledge funding from IARPA under the MQCO program, from NSF under the Materials World Network program, and from the Homing Programme of the Foundation for Polish Science supported by
the EEA Financial Mechanism ({\L}C).  We thank M.~Biercuk and V.~Dobrovitski for useful discussions.
\end{acknowledgments}


\begin{thebibliography}{29}
\bibitem[{CPM()}]{CPMGa2to4params}
\bibinfo{note}{CPMG visibilities for $\alpha=2$ are: $\nn=2, V=0.52\pm0.02$;
  $\nn=4, V=0.46\pm0.04$; $\nn=8, V=0.62\pm0.02$. Visibilities for $\alpha=3$
  are: $\nn=2, V=0.48\pm0.03$; $\nn=4, V=0.43\pm0.02$; $\nn=8, V=0.58\pm0.03$.
  Visibilities for $\alpha=4$ are: $\nn=2, V=0.46\pm0.03$; $\nn=4,
  V=0.42\pm0.02$; $\nn=8, V=0.56\pm0.03$.}
\end{thebibliography}
\end{document}